\input harvmac.tex
\Title{\vbox{\baselineskip12pt\hbox{MRI-PHYS/04/96}
\hbox{CTP-TAMU-02/96}\hbox{hep-th/9601118}}}
{\vbox{\centerline{Rotating Dyonic Black Holes
in Heterotic String Theory}}}
\centerline{\bf Dileep P. Jatkar$^1$, Sudipta Mukherji$^2$ and  Sudhakar
Panda$^1$}
\smallskip\centerline{$^1$\it Mehta Research Institute of Mathematics and
Mathematical Physics}
\smallskip\centerline{\it10, Kasturba Gandhi Marg, Allahabad 211 002, INDIA}
\smallskip\centerline{e-mail: dileep, panda@mri.ernet.in}
\smallskip\centerline{$^2$\it Center for Theoretical Physics, Department of 
Physics}
\smallskip\centerline{\it Texas A \& M University, College Station, Texas
77843-4242, USA}
\smallskip\centerline{e-mail: mukherji@bose.tamu.edu}
\vskip .3in
We study a class of rotating dyonic black holes in the heterotic string
theory in four dimension which have left, right independent electric
charges but have same magnitude for the left and right magnetic charges.
In both left and right sector the electric and the magnetic vectors are
orthogonal to each other. The gyromagnetic (electric) ratios are in general 
found not to have an upper bound.
\Date{1/96}

\def \sha {{ \sinh}\alpha}
\def \cha{{\cosh}\alpha}
\def \shg{{\sinh}\gamma}
\def \chg{{\cosh}\gamma}
\def \sh{{\sinh}}
\def \ch{{\cosh}}
\def \r{\rho}
 \def \c{\cos}
\def\s {{\sin}}
\def\p {\phi}
 \def \t {\theta}
\lref\jmp{D. Jatkar, S. Mukherji and S. Panda, {\it Dyonic 
Black Holes in Heterotic String Theory}, hep-th/9512157.}

\lref\sena{A. Sen, {\it Black Hole Solutions in Heterotic String
Theory on A Torus}, Nucl. Phys. {\bf B440} (1995) 421, hep-th/9411187.}

\lref\youm{M. Cvetic and D. Youm, {\it All the Static Sperically
Symmetric Black Holes of Heterotic String Theory}, hep-th/9512127.}

\lref\rasheed{D. Rasheed, {\it The Rotating Dyonic Black Holes of 
Kaluza-Klein Theory}, Nuc. Phys {\bf B454} (1995) 379.}

\lref\ashoke{A. Sen, {\it Strong-Weak Coupling Duality in
Three Dimensional String Theory}, hep-th/9408083.} 

\lref\youma{M. Cvetic and D. Youm, {\it Dyonic BPS saturated Black Holes
as Exact Superstring solutions}, hep-th/9507090.}

\lref\horn{J. Horne and G. Horowitz, {\it Rotating Dilatonic Black Holes},
Phys. Rev. {\bf D46} (1992) 1340.}

\lref\kall{R. Kallosh, D. Kastor, T. Ortin and T. Torma, {\it Supersymmetry
and Stationary solutions in Dilaton Axion Gravity}, hep-th/9406059; M. Cvetic
and D. Youm, hep-th/9603147.}

\lref\susskind{L. Susskind, {\it Some speculations about black hole entropy
in String theory}, hep-th/9309145; J. Russo and L. Susskind, {\it Asymptotic
level density in heterotic string theory and rotating black holes},
hep-th/9405117; L. Susskind and J. Uglum, Phys. Rev {\bf D50} (1994) 2700.}


\newsec{Introduction}

In a recent paper \refs{\jmp}, we analyzed dyonic black holes
in heterotic string theory on a six torus. In particular,
 we discussed the local as well global structure of  
 those dyonic black holes which are spherically symmetric\foot
{The spherically symmetric case has also been analyzed in \refs{\youm}.}.
It was found that a generic solution of this class depends
on four boost parameters which are responsible for
left, right electric and magnetic charges. Also it depends
on fifty two rotation parameters which set the directions
of electric and magnetic charge vectors. When we allow
these black holes to have angular momentum, the resulting
situation becomes more complicated. Even though, in principle, the local
structure of the metric as well as of the gauge fields and scalars
can be found from our general discussion \refs{\jmp},
we only determined mass, angular momentum and charges looking at the
asymptotic behavior of the configuration. This, in turn, 
allowed us to figure out the cases when the black holes
are extremal, BPS saturated and supersymmetric.

In this letter we find and analyse a particular class of rotating dyonic
black holes, for which we explicitly give the metric and various
other fields. This class of dyons are charecterised by independent
left, right electric charges but they have same magnitude
of the left and 
right magnetic charges. Among other things, we find they have
non-zero gyromagnetic and gyroelectric ratios. However, unlike
the pure electrically charged black holes and black holes
in pure Einstein-Maxwell theory~\refs{\sena,\horn}, 
these ratios {\it do not} have upper bound $2$. They can be 
arbitrarily large.    
We discuss how to arrive at the soultions and study their properties in 
the next section.

\newsec{Rotating Dyonic Black Holes}

It is well known that as long as we consider static configurations
of four dimensional heterotic string theory on six torus, the 
T-duality group and the electric-magnetic duality group of theory
combines in a non-trivial way resulting an $O(8, 24)$ symmetry
of the theory \refs{\ashoke}. It is also known \refs{\sena} that the coset 
$(O(22,2)\times O(6,2))/(O(22)\times O(6) \times SO(2))$ of the 
original $O(8, 24)$ acts as a solution generating group. Namely,
the action of the coset on a known solution can be used to 
generate non-trivial field configuration of the theory. In \refs{\jmp}
we discussed, in detail, how to parametrize the coset. The
explicit procedure for construction of {\it all} rotating black holes
in heterotic string theory on six torus, for fixed asymptotic
configuration of various fields representing asymptotically flat space
time, was also spelled out there, though, as mentioned in the 
introduction, final configurations were only worked out in the
case of static black holes. In what follows, we will explicitly
construct a class of rotating dyonic black holes and discuss
their properties. 

We  start with the Kerr metric
\eqn\kerr{\eqalign{ds^2 = &-{\r^2 + a^2\c^2\t -2m\r\over{\r^2 + a^2\c^2\t}}
dt^2 + {\r^2 + a^2\c^2\t\over{\r^2 + a^2 -2m\r}}d\r^2
+ (\r^2 + a^2\c^2\t)d\t^2\cr
&+ {\s^2\t\over{\r^2 + a^2 \c^2\t}}[(\r^2 + a^2)(\r^2 + a^2 \c^2\t)
+ 2m\r a^2\s^2\t ]d\p^2\cr
&-{4m\r a\s^2\t\over {\r^2 + a^2\c^2\t}}dtd\p ,\cr
&\Phi =0, ~~~~B_{\mu\nu} = 0, ~~~~A_{\mu}^{(a)} = 0, ~~~~M = I_{28},\cr}}
and apply a properly chosen solution generating matrix belonging to 
the coset discussed above
to get
a new rotating black hole geometry. Note that the above configuration is
guaranteed to be a solution of the equations of motion 
of heterotic string theory on six torus
since these equations of motion turn out to be identical to
Einstien equation in the matter free space. We will not repeat the
procedure 
here but only mention that the matrix which generate
rotating dyonic black holes with independent $28$ electric and $28$
magnetic charges are parametrised, among others,  
by four boost parameters that we 
denoted by $\alpha, \beta, \gamma$ and $\delta$ and also rotation
parameters $R$, $T$ and $u$.\foot{The most general form of the 
matrix was given in \refs{\jmp}}
The explicit class of configuration that we generate here are the
one which can be obtained by boosting the above Kerr solution where
the boost parameters are $\alpha$, $\gamma$ and $\beta (=\delta)$.
We also set the rotation parameters $R$, $T$ and $u$ to zero. It is easy
to check that this choice of parameters automatically satisfies the Taub-NUT 
condition given in \refs{\jmp}, 
and hence guarantees the asymptotically flat metric for
the new configuration. Following the procedure of \refs{\sena,\jmp},
after long algebraic manipulations, we get the
new metric in the following form:
\eqn\newmet{\eqalign{&ds^2 = \tilde R\{-\Delta^{-1}Rdt^2
+ (\r^2 + a^2 -2m\r )^{-1}d\r^2 + d\theta^2 + \Delta^{-1}\s^2\theta[\Delta\cr 
&+ a^2 \s^2\theta\{\tilde R + 2m\r \cha\chg - m^2\sh^2\beta
(\cha -\chg )^2\}]d\p^2\cr
&-2\Delta^{-1}m\r a \s^2\theta \ch\beta(\cha + \chg)dtd\p\},}}
where 
\eqn\del{\eqalign{\Delta =& \tilde R(R+2m\r\ch\alpha\ch\gamma+m^2(\ch\alpha-
\ch\gamma)^2)\cr &-m^2a^2\c^2\theta\ch^2\beta(\ch\alpha-\ch\gamma)^2\cr
R &= \r^2 -2m\r + a^2\c^2\theta\quad \tilde R = \r^2 +2m\r\sh^2\beta +
a^2\c^2\theta.}}
Similarly we can extract the exact expressions for various fields in the
transformed solutions. The dilaton is given by
\eqn\dilaton{\phi = {1\over 2} {\rm log}{\tilde R^2\over \Delta}.}
The time components of the gauge fields are ( with $1\leq n \leq 22$ and
$23 \leq p \leq 28$)
\eqn\at{\eqalign{A_t^{(n)} &={R_{22}\over\sqrt{2}}{1\over\Delta}
\pmatrix{0_{20}\cr \cr -ma\c\theta\sh\beta
(\tilde R\ch\gamma\cr +m\r\ch^2\beta(\ch\alpha -\ch\gamma))\cr \cr m\sh\alpha\{
ma^2\c^2\theta\ch^2\beta(\ch\alpha-\ch\gamma)\cr
-\tilde R(m(\ch\alpha-\ch\gamma)+\r\ch\gamma)\}}\cr
A_t^{(p)} &= {R_6\over\sqrt{2}}{1\over\Delta}\pmatrix{0_{4}\cr \cr ma\c\theta
\sh\beta(\tilde R\ch\alpha\cr -m\r\ch^2\beta(\ch\alpha-\ch\gamma))\cr \cr 
m\sh\gamma\{\tilde R(m(\ch\alpha-\ch\gamma)-\r\ch\alpha)\cr -ma^2\c^2\theta
\ch^2\beta(\ch\alpha-\ch\gamma)\}}}}
whereas the spatial components of the gauge fields are as follows
\eqn\aphi{\eqalign{A_{\phi}^{(n)} &={R_{22}\over{2\sqrt{2}}}{1\over\Delta}
\pmatrix{0_{20}\cr \cr m\c\theta\sh 2\beta[\tilde R\{\r^2+a^2
+2m\r(\ch\alpha\ch\gamma-1)\}\cr+m^2(\ch\alpha-\ch\gamma)^2\{\r^2+(2m\r-a^2)
\sh\beta\}\cr +m\r a^2\s^2\theta(\ch\alpha\ch\gamma-\ch^2\gamma)]\cr \cr
2m\r a\sh\alpha\ch\beta\s^2\theta[\tilde R+m(\ch\alpha\ch\gamma-\ch^2\gamma)\cr
\{\r+2m\sh^2\beta\}]}\cr
A_{\phi}^{(p)} &= {R_6\over{2\sqrt{2}}}{1\over\Delta}\pmatrix{0_{4}\cr \cr
-m\c\theta\sh 2\beta[\tilde R\{\r^2+a^2
+2m\r(\ch\alpha\ch\gamma-1)\}\cr+m^2(\ch\alpha-\ch\gamma)^2\{\r^2+(2m\r-a^2)
\sh\beta\}\cr +m\r a^2\s^2\theta(\ch\alpha\ch\gamma-\ch^2\alpha)]\cr \cr 
2m\r a\sh\gamma\ch\beta \s^2\theta[\tilde R+m(\ch\alpha\ch\gamma-\ch^2\alpha)\cr
\{\r+2m\sh^2\beta\}]}}}
where $R_{22}$ and $R_6$ are general $22\times 22$ and $6\times 6$ rotation
matrices.
The scalar fields are given by
\eqn\scalar{M_{ab} = I_{28}+\pmatrix{R_{22}{\cal P}R_{22}^T& R_{22}{\cal Q}
R_6^T\cr (R_{22}{\cal Q}R_6^T)^T& R_6{\cal R}R_6^T}}
${\cal P}$, ${\cal Q}$ and ${\cal R}$ can be written as follows
\eqn\pqr{{\cal P}=\pmatrix{0_{20\times 20}&0&0\cr 0&p_1&p_2\cr 0&p_2&p_3}
\quad{\cal Q}=\pmatrix{0_{4\times 20}&0&0\cr 0&q_1&q_2\cr 0&q_3&q_4}
\quad{\cal R}=\pmatrix{0_{4\times 4}&0&0\cr 0&r_1&r_2\cr 0&r_2&r_3}}
In the above we have defined the following notations:
\eqn\pp{\eqalign{p_1 = &{2m^2\sh^2\beta\over\Delta}[\sh^2\beta\{\r^2-2m\r
(1-\ch\alpha\ch\gamma)\cr
&m^2(\ch\alpha-\ch\gamma)^2\}+\sh^2\gamma a^2\c^2\theta]\cr
p_2 = &{-2m^2a\c\theta\sh\alpha\sh\beta\over\Delta}[\r(\ch^2\beta-\ch^2\gamma)
+m\sh^2\beta(\ch\alpha-\ch\gamma)\ch\gamma]\cr
p_3 = &{2m^2\sh^2\alpha\over\Delta}[\r^2\sh^2\gamma+\sh^2\beta(2m\r\sh^2\gamma
+a^2\c^2\theta)]}}
\eqn\qr{\eqalign{q_1 = &{-2m\sh^2\beta\over\Delta}[\sh^2\beta\{m\r^2-2m^2\r(1-
\ch\alpha\ch\gamma)\cr &+m^3(\ch\alpha-\ch\gamma)^2\}
+\r^3-ma^2\c^2\theta(1+\ch\alpha\ch\gamma)\cr &+\r a^2\c^2\theta
+m^2\r(\ch\alpha-\ch\gamma)^2-2m\r^2(1-\ch\alpha\ch\gamma)]\cr
q_2 = &{2ma\c\theta\sh\beta\sh\gamma\over\Delta}[\r^2-m\r(1-\sh^2\beta-
\ch\alpha\ch\gamma)\cr
&+a^2\c^2\theta-m^2\ch\alpha(\ch\alpha-\ch\gamma)\sh^2\beta]\cr
q_3 = &{2ma\c\theta\sh\alpha\sh\beta\over\Delta}[\r^2+m\r(\sh^2\beta+
\ch\alpha\ch\gamma)\cr
&a^2\c^2\theta+m^2\sh^2\beta\ch\gamma(\ch\alpha-\ch\gamma)]\cr
q_4 = &{-2m\sh\alpha\sh\gamma\over\Delta}[\r^3+m\r^2(\ch\alpha\ch\gamma-1+
2\sh^2\beta)\cr
&+2m^2\r\sh^2\beta(\ch\alpha\ch\gamma-1)+a^2\c^2\theta(\r+m\sh^2\beta)]\cr
r_1 = &{2m^2\sh^2\beta\over\Delta}[\sh^2\beta\{ \r^2-2m\r(1-\ch\alpha\ch\gamma)
+m^2(\ch\alpha-\ch\gamma)^2\}\cr
&+\sh^2\alpha a^2\c^2\theta]\cr
r_2 = &{-2m^2a\c\theta\sh\beta\sh\gamma\over\Delta}[\r(\ch^2\alpha-
\ch^2\beta)\cr
&+m\sh^2\beta\ch\alpha(\ch\alpha-\ch\gamma)]\cr
r_3 = &{2m^2\sh^2\alpha\over\Delta}[\r^2\sh^2\alpha+(2m\r\sh^2\alpha
+a^2\c^2\theta)\sh^2\beta]}.}
The antisymmetric tensor field is given by
\eqn\ast{\eqalign{B_{t\phi}&={\ch\beta(\ch\alpha-\ch\gamma)\over\Delta}
[m\r a\s^2\theta\{\tilde R +m(\ch\alpha\ch\gamma-1)(\r+2m\sh^2\beta)\}\cr
&-\sh^2\beta m^2a\c^2\theta(\r^2+a^2-2m\r)].}}

To get various charges associated with  the black hole, we rewrite 
the metric in the Einstein frame
\eqn\enewmet{\eqalign{ds_{E}^2 = &e^{-\Phi} ds^2\cr = &{\sqrt \Delta}
\{-\Delta^{-1}(\r^2 -2m\r + a^2\c^2\theta)dt^2\cr
&+ (\r^2 + a^2 -2m\r )^{-1}d\r^2 + d\theta^2\cr
&+ \Delta^{-1}\s^2\theta[\Delta + a^2 \s^2\theta
\{\r^2 + a^2\c^2\theta + 2m\r \cha\chg\cr 
&+ m(\ch^2\beta -1)(2\r
-m(\cha -\chg )^2)\}]d\p^2\cr
&-2\Delta^{-1}m\r a \s^2\theta \ch\beta(\cha + \chg)dtd\p\}}}

This metric clearly corresponds to a class of rotating 
black holes characterized by different values of boost parameters
$\alpha, \beta$ and $\gamma$. The mass $M$, angular momentum $J$,
electric charges $Q^{L,R}_{elc}$, 
magnetic charges $Q^{L,R}_{mag}$,
electric and magnetic dipole moments $\mu^{L,R}_{elc}$ and $\mu^{L,R}_{mag}$
are found to be as follows
\eqn\mass{M = {1\over2}m(\ch^2\beta + \cha\chg)}
\eqn\ang{J = {1\over 2} ma \ch\beta (\cha + \chg)}
\eqn\elecch{ Q^{L}_{elc}= {m\over {\sqrt 2}}\sha\chg R_{22}\pmatrix{0_{20}\cr
0\cr 1},
~~~Q^{R}_{elc}= {m\over {\sqrt 2}}\shg\cha R_{6}\pmatrix{0_{4}\cr 0\cr 1} .}
\eqn\magch{Q^{L}_{mag}=  {m\over {2\sqrt 2}}\sh 2\beta R_{22}\pmatrix{0_{20}\cr
1\cr 0},
~~~Q^{R}_{mag} =-{m\over {2\sqrt 2}}\sh 2\beta R_{6}\pmatrix{0_4\cr 1\cr 0} .}
\eqn\elecmom{\mu^{L}_{elc}= {ma\over {\sqrt 2}}\sh\beta\chg
R_{22}\pmatrix{0_{20}\cr 1\cr 0},
~~~\mu^{R}_{elc} =-{ma\over {\sqrt 2}}\sh\beta\cha R_6\pmatrix{0_4\cr 1\cr 0}.}
\eqn\magmom{\mu^{L}_{mag}= {ma\over {\sqrt 2}}\sha\ch\beta
R_{22}\pmatrix{0_{20}\cr 0\cr 1},
~~~\mu^{R}_{mag}= {ma\over {\sqrt 2}}\shg\ch\beta R_6\pmatrix{0_4\cr 0\cr 1}.}
The superscripts $L, R$ denote the left and right sectors respectively which
are defined by $A^{L(R)}_a={1\over 2} (I_{28}-(+)L)_{ab}A_b$ for a vector $A$.
The necesity of defining these vectors for left and right sectors
separately is that for generic values 
of parameters $\alpha, \beta$ and $\gamma$, the
electric(magnetic) charge vectors are {\it not} parallel
to each other and thus we will not be able to define the gyromagnetic
(electric) ratios in general. We can apply various  
consistency checks on our solution \newmet\ -~\magmom\ . For example, if we 
set the parameter $a$ to zero, we get the spherically symmetric cofiguration 
discussed
in \refs{\jmp}, on the other hand, setting $\beta$ to zero,
we get the purely electrically charged rotating black holes
discussed in \refs{\sena}.  

Since we have defined the charges and the dipole moments in the left and
right sectors separately we are in a position to evaluate the gyromagnetic
($g_{mag}$) and the gyroelectric ($g_{elc}$) ratios for the left and right 
hand sectors independently. They are found to be
\eqn\gyromag{\eqalign{&g^L_{mag}
= {2 \mu^L_{mag} M\over Q^L_{elc}J} = 2{\cha\chg + \ch^2\beta\over{(
\cha + \chg)\chg}},\cr
&g^R_{mag} 
={2\mu^R_{mag}M\over Q^R_{elc}J} = 2{ \cha\chg + \ch^2\beta\over{(
\cha + \chg)\cha}}}}
\eqn\gyroelec{\eqalign{&g^L_{elc} 
= {2\mu^L_{elc}M\over Q^L_{mag}J} = 2{(\cha\chg + \ch^2\beta)\chg\over{(
\cha + \chg)\ch^2\beta}},\cr
&g^R_{elc}  
= {2\mu^R_{elc}M\over Q^R_{mag}J} = 2{(\cha\chg + \ch^2\beta)\cha
\over{(
\cha + \chg)\ch^2\beta}}}}
Here the superscript on $g$ denotes the handedness 
as before. 
In the case of
pure electrically charged black holes, as was noticed earlier
\refs{\horn, \sena}, the 
gyromagnetic ratios have upper bound $2$. Same is true for 
black holes in pure Einstein-Maxwell theory. On the other hand,
in our case, the value of the gyromagnetic(electric) ratios
{\it do not} seem to have any upper bound. 
Similar phenomenon was also noticed recently in \refs{\rasheed} for the
dyonic black holes in the Kaluza-Klein theory\foot{However
we differ from \refs{\rasheed} 
in the following point. In our case by construction
the electric and magnetic vectors are orthogonal. \refs{\rasheed}
on the other hand has only one $U(1)$ with both electric and magnetic
charge and hence parallel to each other. In a suitable $O(6, 22)$
frame, these charges in our case belong to different $U(1)$'s. Thus 
for dyonic black holes, our results are different from him by
construction. However, if we restrict to either electrically charged
or magnetically charged black holes, we can compare the results. For example, 
both reduces to Sen's result of purely electrically charged 
black holes. If, on the other hand, we restrict ourself only to the 
magnetically charged black hole ($\alpha = 
\gamma = 0$ in our case), we find that the magnetically
charged solution of \refs{\rasheed} ($\alpha = 0$ in his)  
and our's are the same after appropriate identifications of coordinates.}
To see it more 
explictly let us study the following cases:
\item{(1)} $\beta = \gamma$
\eqn\limita{g^L_{elc} = g^L_{mag} =2,~~~g^R_{elc} = 2{\ch\alpha\over
\ch\beta}, g^R_{mag} = 2{\ch\beta\over\ch\alpha}}
\item{(2)}$\alpha = \gamma$
\eqn\limitb{g^{L,R}_{elc} = 1 + {\ch^2\alpha\over \ch^2\beta},~~~
g^{L,R}_{mag} = 1 + {\ch^2\beta\over \ch^2\alpha}}

\noindent In the first case, we find that the left handed gyromagnetic
(electric) ratios are equal to $2$ but right handed gyromagnetic(electric) 
ratios can increase (decrease) as $\beta$ ($ > \alpha$)
increases and vice versa when $\alpha$ ($ > \beta$) increases. In other
words, the right handed gyromagnetic(electric) 
ratio effectively depends on the charge 
ratio ${|Q^R_{mag}|\over |Q^R_{elc}|}({|Q^R_{elc}|\over |Q^R_{mag}|})$.
Whenever one of the charge ratios gets large, gyromagnetic(or electric) ratio
grows without bound. 
The chiralities will be exchanged if we replace $\gamma$ by $\alpha$.
In the second case however we find that the result is independent
of chiralities but depends upon electric and magnetic nature of
the ratios. In the limit of large $\alpha$ or $\beta$ either
electric or magnetic ratio grows and the other one approaches $1$.
In other words, for large charges the gyromagnetic(electric) 
ratio behaves as $1 + {|Q_{mag}|\over |Q_{elc}|}(1 + |{Q_{elc}|\over 
|Q_{mag}|})$.
So again, when one of the charge ratios becomes large, gyromagnetic
(or electric) ratios grow without bound.

The singularity structure of the new solution remains the same as
the original Kerr metric, i.e., the singularity apears
at $\r = 0$, and the inner and outer horizons are located at 
\eqn\locus{\r = m \pm {\sqrt{m^2 -a^2}}}
In order to avoid naked singularity, we need to have $a < m$.
The extremal limit is reached when $a$ approaches $m$.

Now we will analyse various thermodynamic quantities. One such important
quantity is the area of the event horizon as it determines the
classical Bekenstein-Hawking entropy. We find
\eqn\area{ A = 4\pi m \ch\beta(\cha + \chg)( m + {\sqrt{m^2 -a^2}}). }
So the entropy is
\eqn\entropy{S = {A\over 4} = \pi m\ch\beta(\cha + \chg)
(m + {\sqrt{m^2 -a^2}}).}
The
surface gravity which in our case is
\eqn\sgra{\kappa = {{\sqrt{m^2 -a^2}}\over {m\ch\beta
(\cha + \chg)( m + {\sqrt{m^2 -a^2}})}}.}
Hence the tempreature of the black hole is
\eqn\temp{T = {\kappa\over 2\pi} = {{\sqrt{m^2 - a^2}}\over 
{2 m \pi\ch\beta (\cha + \chg)(m + {\sqrt{m^2 -a^2}})}}.}

Comparing \ang\ and \entropy\ , we see that the entropy increases 
(decreases) with the increase (decrease) of angular momentum.
On the other hand, the temperature decreases with the increase of the
angular momentum. However, the product of entropy and temperature
is independent of the parameters $\alpha$, $\beta$ and $\gamma$
and goes to zero in the extremal limit $a \rightarrow m$.
Also the temperature  vanishes in this extremel limit. This is quite
similar to the case of purely electrically charged holes \refs{\sena}
\foot{However, for
the dyonic black holes with unconstrained left, right magnetic
charges, entropy {\it is finite} for spherically symmetric 
case as can be seen
in \refs{\youma, \jmp}}. A little analysis shows that for fixed $m$ 
and $a$, the entropy and temperature can be expressed in terms of 
the electric and magnetic charges. However the expressions are not very 
illuminating so we do not display it here. But relations simplify as the
charges become large and in this limit entropy always increases with the 
charges. 

There are several limits of this class of solutions where 
it becomes supersymmetric and BPS saturated. But in all these 
cases, one has to set the rotation parameter $a$ to zero otherwise the
horizons disappear leaving behind a naked singularity. This fact has been
observed earlier also\refs{\kall}. Therefore the supersymmetric
solution reduces to static spherically symmetric dyonic black hole.
Analysis of these black holes have already been presented in our
earlier paper \refs{\jmp}. We expect that our results will be useful to
get a better understanding of the relationship between black holes and
elementary string states in the same spirit of refs.\refs{\susskind}.

{\bf Acknowledgement}: We would like to thank A. Sen for discussions. Work of
SM is supported by NSF Grant No. PHY-9411543.
\vfill\eject
\listrefs
\bye